\documentclass[10pt,conference,letterpaper]{IEEEtran}

\usepackage{amsmath,amssymb}
\usepackage{graphicx}
\usepackage{multirow}
\usepackage{cite}
\usepackage[colorlinks=true,linkcolor=black,citecolor=black,pdfborder={0 0 0}]{hyperref}
\usepackage{enumitem} % 控制列表间距
\setlist[itemize]{itemsep=0pt, topsep=2pt}
\setlist[enumerate]{itemsep=0pt, topsep=2pt}
\usepackage{microtype}
\usepackage{algorithm}
\usepackage{algpseudocode}
\usepackage{threeparttable}
\usepackage{xspace}
\usepackage{pifont}
\usepackage{pdfpages}
\usepackage{booktabs}
\usepackage{float}
\usepackage{subcaption} % 用于子图

\usepackage{perpage}
\usepackage{footmisc}

\begin{document}

% \title{\huge Toward Adaptive Disk Failure Prediction via Stream Mining}

% \author{Shujie Han$^1$, Patrick P. C. Lee$^1$, Zhirong Shen$^2$, 
% Cheng He$^3$, Yi Liu$^3$, and Tao Huang$^3$\\
% $^1$The Chinese University of Hong Kong \ \
% $^2$Xiamen University \ \
% $^3$Alibaba Group}

\title{BVLSM: Write-Efficient LSM-Tree Storage via WAL-Time Key-Value Separation\\
}

\author{
\IEEEauthorblockN{
    Ming Li\textsuperscript{1*} \quad
    Wendi Cheng\textsuperscript{1*} \quad
    Jiahe Wei\textsuperscript{1} \quad
    Xueqiang Shan\textsuperscript{1} \quad
    Weikai Liu\textsuperscript{2} \quad
    Xiaonan	Zhao\textsuperscript{1} \quad
    Xiao Zhang\textsuperscript{1$^{\dagger}$} \quad
    \\[5pt]
    % $^1$Northwestern Polytechnical University, Xi’an, China \qquad
    % $^2$Zhongtian Rocket Technology Co.$^1$, Ltd, Xi’an, China
}
\IEEEauthorblockA{$^1$Northwestern Polytechnical University, Xi’an, China}
\IEEEauthorblockA{$^2$Zhongtian Rocket Technology Co., Ltd, Xi’an, China}

\IEEEauthorblockA{
    Email: \{lming,chengwendi,vjiahe,shanxueqiang\}@mail.nwpu.edu.cn,\{zhaoxn,zhangxiao\}@nwpu.edu.cn,wakenliu@sina.com
}
}

\newcommand\blfootnote[1]{%
  \begingroup
  \renewcommand\thefootnote{}\footnote{#1}%
  \addtocounter{footnote}{-1}%
  \endgroup
}

\maketitle

\begin{abstract}
Modern data-intensive applications increasingly store and process big-value items, such as multimedia objects and machine learning embeddings, which exacerbate storage inefficiencies in Log-Structured Merge-Tree (LSM)-based key-value stores. This paper presents BVLSM, a Write-Ahead Log (WAL)-time key-value separation mechanism designed to address three key challenges in LSM-Tree storage systems: write amplification, poor memory utilization, and I/O jitter under big-value workloads. Unlike state-of-the-art approaches that delay key-value separation until the flush stage, leading to redundant data in MemTables and repeated writes. BVLSM proactively decouples keys and values during the WAL phase. The MemTable stores only lightweight metadata, allowing multi-queue parallel store for big value. The benchmark results show that BVLSM significantly outperforms both RocksDB and BlobDB under 64KB random write workloads. In asynchronous WAL mode, it achieves throughput improvements of 7.6× over RocksDB and 1.9× over BlobDB.
\end{abstract}

\blfootnote{$^*$Equal contribution. $^\dagger$Corresponding author: Xiao Zhang.}

\section{Introduction}
Key-value (KV) storage serves as a fundamental component in modern storage architectures, playing a vital role in write-intensive scenarios such as distributed databases, log-analysis, and caching systems. To improve write efficiency, many high-throughput KV engines use the Log-Structured Merge Tree (LSM-Tree) data structure. Representative LSM-Tree–based storage systems include LevelDB~\cite{leveldb}, RocksDB~\cite{rocksdb}, HBase~\cite{vora2011hadoop}, Cassandra~\cite{lakshman2010cassandra}, and TiDB~\cite{pingcap2024tidb}. These systems are widely deployed in real-time data processing, stream analytics, and graphics workloads, and have become the standard for write-intensive storage engines.

% To improve write efficiency, many high-throughput KV engines adopt the \textit{Log-Structured Merge Tree} (LSM-Tree) data structure, which transforms random writes into sequential appends to reduce disk seek overhead. The core idea is to maximize disk throughput through sequential append writing (rather than in-place updating), while maintaining data orderliness through a multi-level merging (Compaction) mechanism in the background.

\begin{figure*}[t]
    \centering
    \begin{subfigure}{0.48\textwidth}
        \centering
        \includegraphics[width=\linewidth]{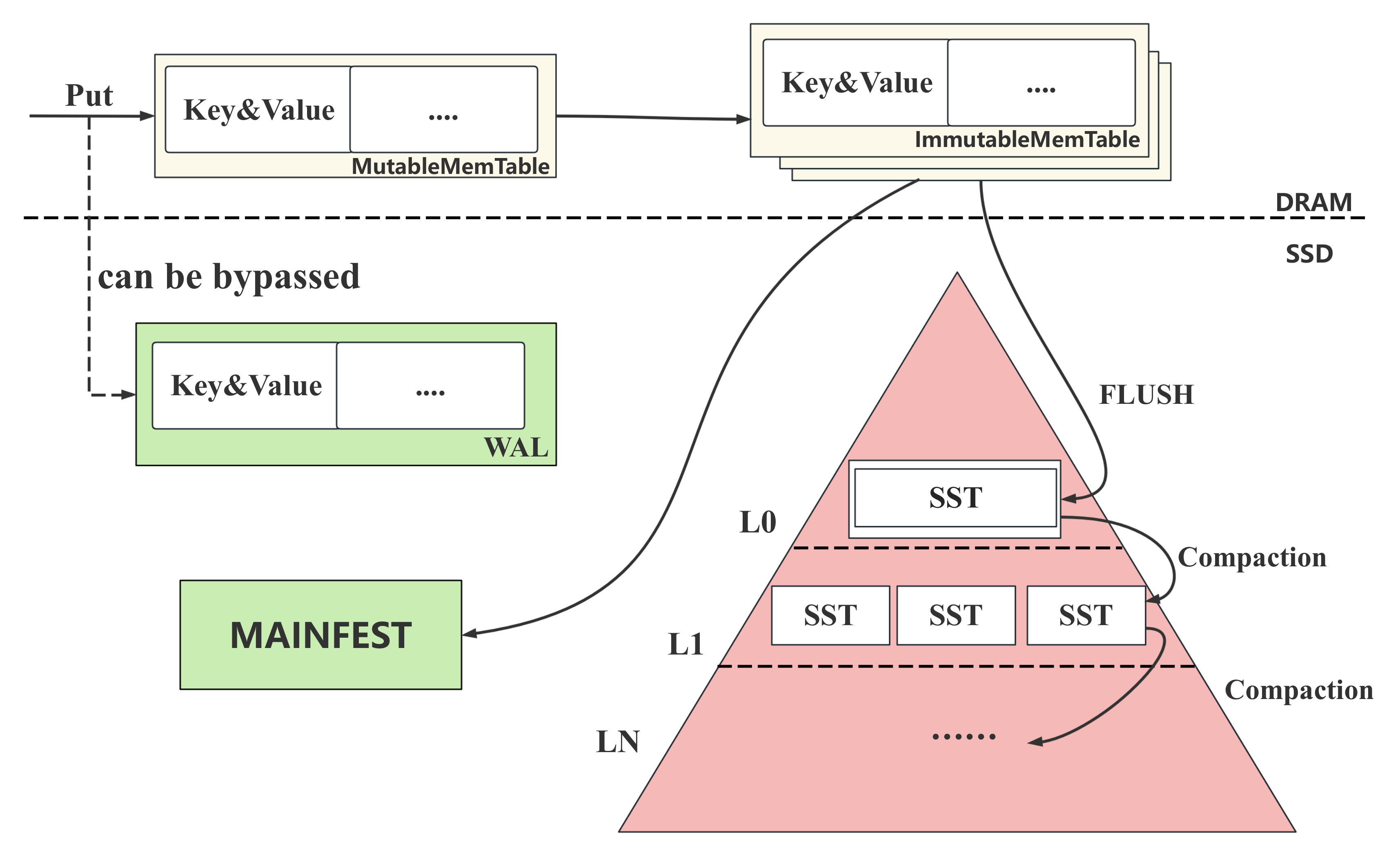}
        \caption{Traditional LSM-Tree architecture}
    \end{subfigure}
    \hfill
    \begin{subfigure}{0.48\textwidth}
        \centering
        \includegraphics[width=\linewidth]{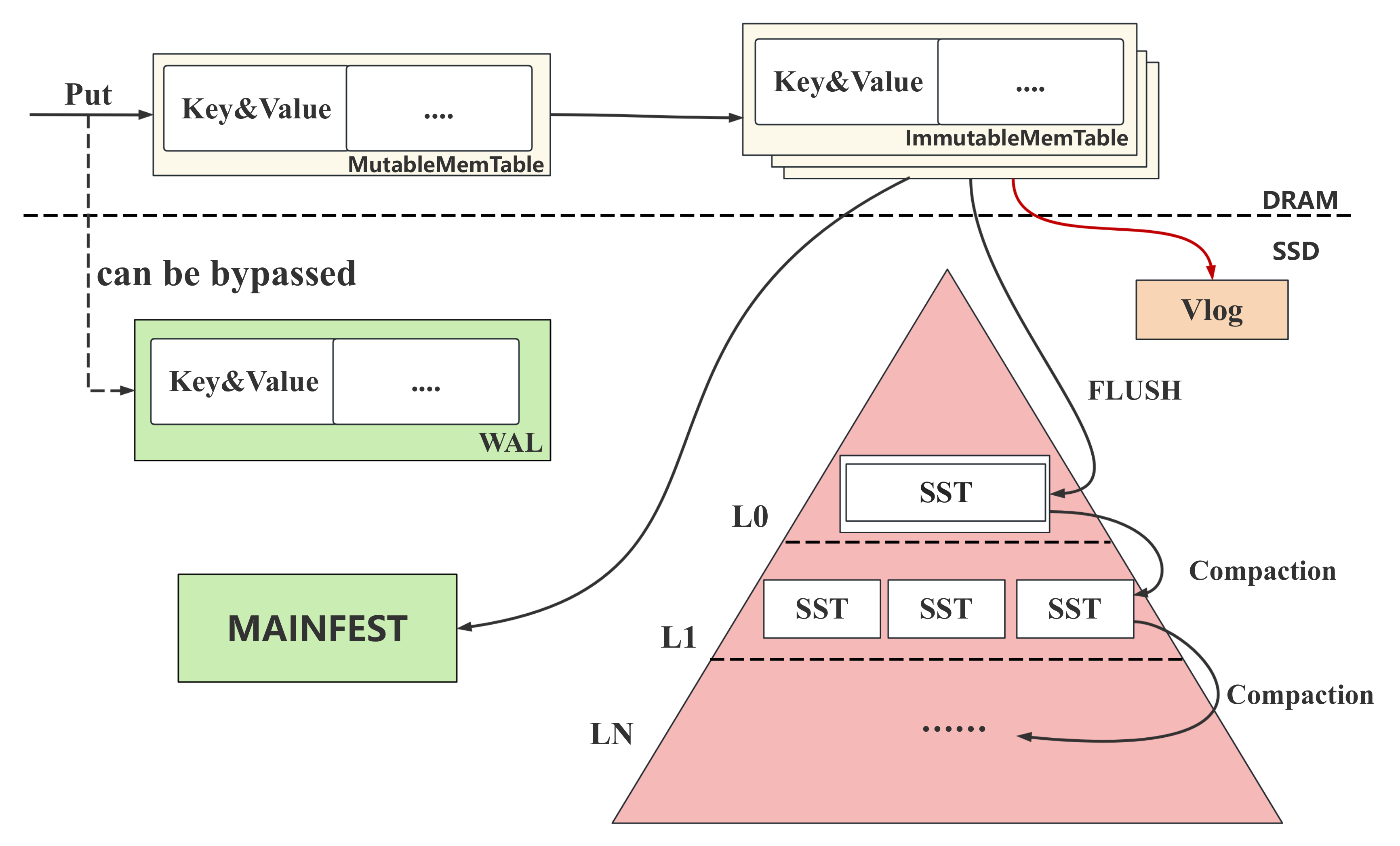}
        \caption{KV-Separated LSM-Tree architecture}
    \end{subfigure}
    \caption{Traditional vs. KV-Separated LSM-Tree Architectures.}
    \label{fig:LSM-Tree_arc}
\end{figure*}

The LSM-Tree is a hierarchical, ordered, and disk-oriented data structure designed for write-intensive workloads~\cite{oneil1996lsm}. By buffering updates in memory and appending them sequentially to disk, the LSM-Tree transforms random writes into large sequential writes, thereby minimizing disk seek time. As illustrated in Figure~\ref{fig:LSM-Tree_arc}\,(a), the \textsc{PUT} request is first appended to a \emph{write-ahead log} (WAL), which can be optionally bypassed for performance, and then inserted into a memory structure called \emph{MemTable}. Only \emph{MutableMemTable} is active at any time, while previously filled ones become \emph{ImmutableMemTables}. When the MemTable reaches a predefined size threshold, it is \textit{flushed} to disk as an immutable \emph{sorted string table(SSTable)}. As the capacity of \emph{SSTable} grows, a background \textit{compaction} process is periodically triggered to merge multiple \emph{SSTables} into a larger \emph{SSTable} at the next level.

However, both the \textit{flush} and \textit{compaction} processes essentially involve rewriting key-value data. This becomes particularly problematic in scenarios with big-size values, where the traditional key-value coupled storage architecture causes value data to be repeatedly migrated across levels. As a result, a significant \textit{write amplification} effect is introduced~\cite{sun2019exploring}, which negatively impacts the overall efficiency of system.

To mitigate \textit{write amplification} in LSM-Tree-based storage systems, key-value separation (KVS) has emerged as a widely adopted optimization strategy~\cite{luo2016wisckey,yao2020matrixkv}. Figure~\ref{fig:LSM-Tree_arc} (b) shows the architecture of a typical KVS LSM-Tree. Keys are stored in an LSM-Tree while values are stored in a separate value-log file(vLog). The artificial value stored along with the key in the LSM-Tree is the address of the actual value in the vLog. As a result, only the keys need to be rewritten during compaction, significantly reducing unnecessary data rewriting and improving write efficiency.

Although many KVS works have been proposed to sovle the \textit{write amplification} problem in LSM‑Tree, yet each introduces its own trade‑offs. \emph{Tiered compaction}, adopted by systems such as Cassandra~\cite{lakshman2010cassandra} and RocksDB~\cite{rocksdb}, accumulates a larger set of \emph{SSTables} before merging.  
Because far more files are combined in a single step, key–value pairs at the same level are rewritten fewer times, reducing write amplification.  
However, the coexistence of many overlapping \emph{SSTables} increases read amplification and amplifies space overhead. An orthogonal line of work, \emph{flush‑time key‑value separation} exemplified by WiscKey~\cite{luo2016wisckey}—decouples values from keys when a \emph{MemTable} is flushed. Both the key and its value remain in memory until the flush occurs, after which only the key (plus a pointer) is written to the LSM‑Tree, while the value is redirected to an append‑only log.  
Although this design simplifies system logic and avoids rewriting large values during compaction, it places substantial pressure on DRAM, lowers effective \emph{MemTable} capacity, and can trigger longer flush latency, as well as write stalls.

% Existing key-value separation schemes (e.g., WiscKey~\cite{luo2016wisckey}) typically perform key-value decoupling during the flush phase, where both keys and values are retained in the MemTable until they are written to disk. Although this approach simplifies the system architecture, it significantly increases memory pressure, reduces MemTable space efficiency, and may lead to higher flush latency and write stalls. 

To address these challenges, we propose BVLSM, a RocksDB-based LSM-Tree key-value separation mechanism designed to tackle three critical issues in LSM-Tree storage systems: write amplification, poor memory utilization, and I/O jitter under big-value workloads. The key innovations of BVLSM include the following:
\begin{itemize}
    \item \textbf{Early-stage key-value separation}: BVLSM introduces a early-separation mechanism that decouples keys and values during the WAL write phase. Only the key and a pointer to the separated value (e.g., offset and length) are recorded in the write path, effectively reducing the data volume involved in both the flush and compaction stages.
    \item \textbf{Multi-queue value parallel store for big value}: BVLSM leverages multi-queue characteristic to maximize the device performance of NVMe SSD and the nature of big-size values. After key-value separation, multiple large values are dispatched to separate queues and written into a dedicated value log file (BValue) via parallel I/O channels. Compared to traditional single-queue writes, this design improves write throughput by up to 40\%.
    \item \textbf{Big value cache for read optimization}: For read scenarios, BVLSM introduces an independent big-value cache implemented as a fixed-size in-memory deque. It employs a Most Recent Write First (MRWF) replacement policy to temporarily store values that have not yet been persisted, thereby improving read responsiveness for recently written data.
\end{itemize}

\section{Background and Related Work}
In this section, we first introduce LSM-Tree and RocksDB, then discuss the challenges that RocksDB faces in handling big-value workloads, and finally summarize the related work.

\subsection{LSM-Tree and RocksDB}
The Log-Structured Merge Tree (LSM-Tree)~\cite{oneil1996lsm} is a disk-oriented data structure designed for write-intensive workloads. By transforming random writes into sequential operations and employing hierarchical compaction, it significantly enhances write efficiency, making it particularly well-suited for solid-state drives (SSDs).  RocksDB~\cite{rocksdb} is a high performance key value store built on the principles of the LSM-Tree. As illustrated in Figure~\ref{fig:LSM-Tree_arc}, the RocksDB architecture consists of the following key components:

\begin{itemize}[itemsep=0pt, topsep=1pt]
    \item \textbf{MemTable}: The \emph{MemTable} consists of one writable \emph{Mutable MemTable} and multiple \emph{Immutable MemTables}. The buffer is typically implemented as a skip list to support efficient in-memory indexing and lookups.
    \item \textbf{Write-Ahead Log (WAL)}: The \emph{WAL} ensures crash recovery by sequentially logging updates to persistent storage before applying them to memory structures.
    \item \textbf{SSTable (Sorted String Table)}: An \emph{SSTable} is an immutable and sorted file that stores key-value pairs in compressed data blocks. It typically includes index blocks and optional Bloom filters to accelerate key lookups.
    \item \textbf{Manifest}: The \emph{Manifest} is a metadata file that maintains the versioned hierarchy of \emph{SSTables} and tracks key-range mappings across levels to ensure reliability and consistency.
\end{itemize}

In RocksDB, data are organized into a multi-level hierarchy from Level~0 (L0) to Level~$n$ (LN), where the capacity of each level is typically 10 times that of the previous level. Higher levels store more stable data with stronger temporal locality. Because \emph{SSTables} in L$0$ may have overlapping key ranges, RocksDB triggers a multi-way merge compaction to sort these \emph{SSTables} and push the resulting data down to L$1$. When a write request arrives, it is first appended to the write-ahead log (WAL) for durability, then inserted into a mutable \emph{MemTable} that supports in-place updates. Once \emph{MemTable} reaches its size threshold, it is promoted to \emph{Immutable MemTable} and flushed to the L~$0$ level on SSD. Then a new \emph{Mmutable MemTable} is allocated to receive subsequent writes.

RocksDB provides configurable WAL modes to balance durability and performance. In the default \textbf{synchronous mode}, each write is followed by an \texttt{fsync} to persist log records on disk, ensuring strict durability guarantees. In \textbf{asynchronous mode}, WAL writes are buffered in memory and flushed to disk in batches, improving throughput at the cost of potential data loss on crash. WAL can also be bypassed for workloads that tolerate data loss but demand maximum write throughput, in which case only \emph{MemTable} and compaction persistence are relied upon.

\subsection{Challenges of Storing Big Values in LSM-Tree-based Storage Systems}
LSM-Tree-based key-value storage systems typically rely on background threads or manual flushes to trigger compaction, thus reducing read amplification by limiting the count of \emph{SSTable} and space amplification by removing obsolete entries. However, this strategy incurs severe write amplification ---- up to 42$\times$ in RocksDB and 27$\times$ in LevelDB ---- resulting in excessive disk I/O and frequent I/O jitter~\cite{raju2017pebblesdb}. The problem is even more detrimental on SSDs due to their inability to perform in-place updates, erase-before-write requirement, and limited write endurance. During multilevel compaction, the \emph{SSTables} at level $L_{i-1}$ are fully read, merged, and rewritten at level $L_i$. With the standard 10:1 size ratio, compacting one \emph{SSTable} at $L_i$ may involve reading ten \emph{SSTables} from $L_{i-1}$, producing a combined read/write amplification exceeding 10$\times$.

\begin{figure}
    \centering
    \includegraphics[width=1.0\linewidth]{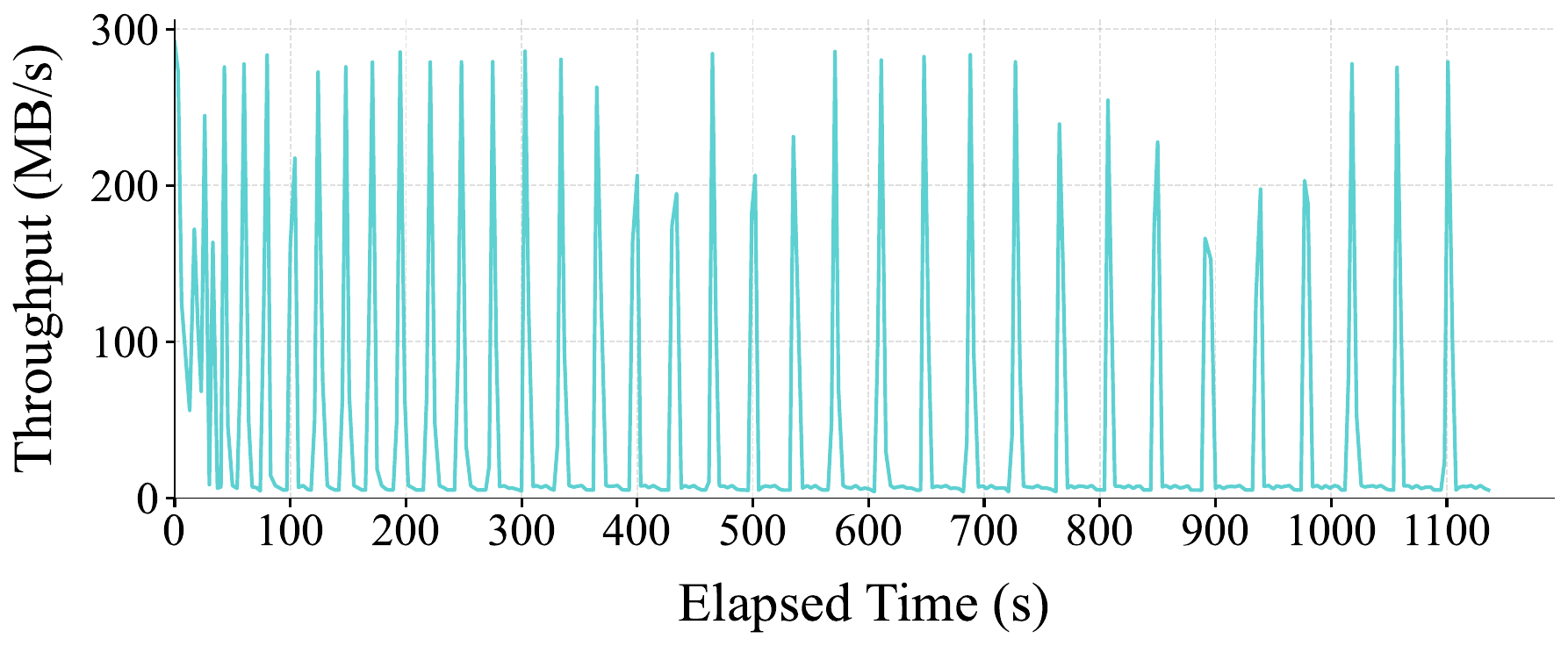}
    \caption{RocksDB instant throughput and average throughput. The blue curve shows the instant throughput every 10 seconds. The red curve shows the cumulative average throughput.}
    \label{fig:RocksDB real-time throughput}
\end{figure}

Moreover, \textbf{I/O jitter} arises when Level~0 becomes saturated. To illustrate this, we examine the runtime behavior of RocksDB under write-intensive workloads. As shown in Figure~\ref{fig:RocksDB real-time throughput}, both instantaneous throughputs exhibit noticeable fluctuations, reflecting the unstable I/O performance of the system. This stagnation occurs when the number of \emph{SSTables} in Level~0 exceeds a predefined threshold, preventing \emph{Immutable MemTables} from being flushed in time and thus stalling incoming writes. The situation is further exacerbated by two key factors: (1) \textit{I/O bandwidth contention} between \emph{MemTables} flushes and background compaction processes, and (2) \textit{limited compaction parallelism} due to overlapping key ranges among L0 \emph{SSTables}. The latter constraint forces the L0-to-L1 compactions to proceed serially to maintain the global sort order in Level~1.

%todo:引用验证

In typical big-scale key-value storage scenarios, the size of the value exhibits a significant long-tail distribution characteristic. For example, in TiDB, the average size of the value in distributed transaction logs reaches 256 KB~\cite{pingcap2024tidb}; In the Atlas cloud storage system, the median size of the value exceeds 128 KB~\cite{lai2015atlas}; and in the Facebook production environment, approximately 30\% of the values are greater than 10 KB~\cite{facebook_blobdb}. However, in the existing LSM-Tree architecture, the performance of \emph{MemTable} is highly susceptible to the size of the value. When the value size is significantly larger than that of the key, the storage efficiency of \emph{MemTable} is significantly reduced. Taking a 64 MB \emph{MemTable} as an example, it can only store about 16,320 KV pairs. If 100 GB of data needs to be written, approximately 1,600 \emph{MemTable} are required for storage. However, the number and size of the mutable and \emph{Immutable MemTable} in the system are usually fixed. This limitation leads to frequent flush operations, which in turn impose continuous I/O pressure on the L0 layer of the LSM-Tree. This competition further exacerbates I/O jitter, thereby causing instability in system performance.

\subsection{Related Work}

To address the challenges of write amplification and I/O jitter in LSM-Tree-based storage engines, WiscKey~\cite{luo2016wisckey} first introduces the KVS mechanism, which decouples the storage of keys and values to improve write performance, especially in workloads with big values. As illustrated in Figure~\ref{fig:LSM-Tree_arc} (b), the original KV pair is split such that the key and a lightweight value reference (e.g. offset and length) are retained in the LSM-Tree, while the full value is redirected to a separate append-only value log~(vLog). For example, with a key size of 16B and a value size of 1KB, a 100GB dataset can reduce the LSM-Tree size to around 2GB after separation (assuming a 12B value reference). This design yields two significant benefits: (i) On the write path, compaction only involves keys and metadata, reducing write amplification by up to two orders of magnitude; (ii) On the read path, the depth of \emph{SSTable} traversal for point queries decreases from an average of five levels to two or three, enabling up to 40\% lower read latency through better cache locality and bloom filter integration.

Existing KVS techniques can be categorized into two groups: \textbf{separation at the Flush stage} and \textbf{separation during Compaction}.

\textbf{Flush-stage separation.} This approach performs the KV separation when flushing \emph{MemTables} to the LSM-Tree L0 tier. Systems such as WiscKey~\cite{luo2016wisckey} and TiDB~\cite{pingcap2024tidb} follow this design. In TiDB, for instance, separating values during flush enables the entire LSM-Tree metadata to reside in memory, reducing memory usage from 120GB to 2.5GB under a 100K QPS OLTP workload. BlobDB~\cite{facebook_blobdb}, an official RocksDB extension, adopts a similar architecture but introduces a dynamic threshold mechanism: values exceeding a configurable size are separated, while smaller values remain inline. HashKV~\cite{chan2018hashkv} also performs flush-time separation and optimizes value management through hash-based partitioning and end-to-end validation.

\textbf{Compaction-stage separation.} Another part of the work performs KV separation during the compaction process by offloading cold or big values. Titan~\cite{pingcap_titan} separates big values to BlobFile during compaction and uses background garbage collection and dynamic thresholding. Kreon~\cite{papagiannis2021kreon} separates cold data during compaction while retaining hot data in the LSM-Tree to preserve access efficiency. DiffKV~\cite{li2021differentiated} further introduces similarity-aware compaction and data deduplication to reduce storage costs. Delta-LSM~\cite{li2024enhancing} combines key-value separation with delta encoding to optimize frequent updates to big-values, reducing compaction overhead by only storing incremental changes.

Although existing key-value separation techniques are effective in reducing write amplification and I/O overhead, they share a critical limitation. The complete key-value pairs must still reside in memory and be written to the WAL to maintain consistency. This design leads to substantial memory pressure, especially under high-throughput or big-value workloads. Consequently, there is a strong need for more memory-efficient separation mechanisms that not only reduce write amplification, but also enhance in-memory utilization in big-value storage scenarios.

\section{Design of BVLSM}
In this section, we present BVLSM, a key-value separation mechanism optimized for LSM-Tree-based systems, and describe its overall architecture and core design components.

\begin{figure}[t]
    \centering
    \includegraphics[width=1.0\linewidth]{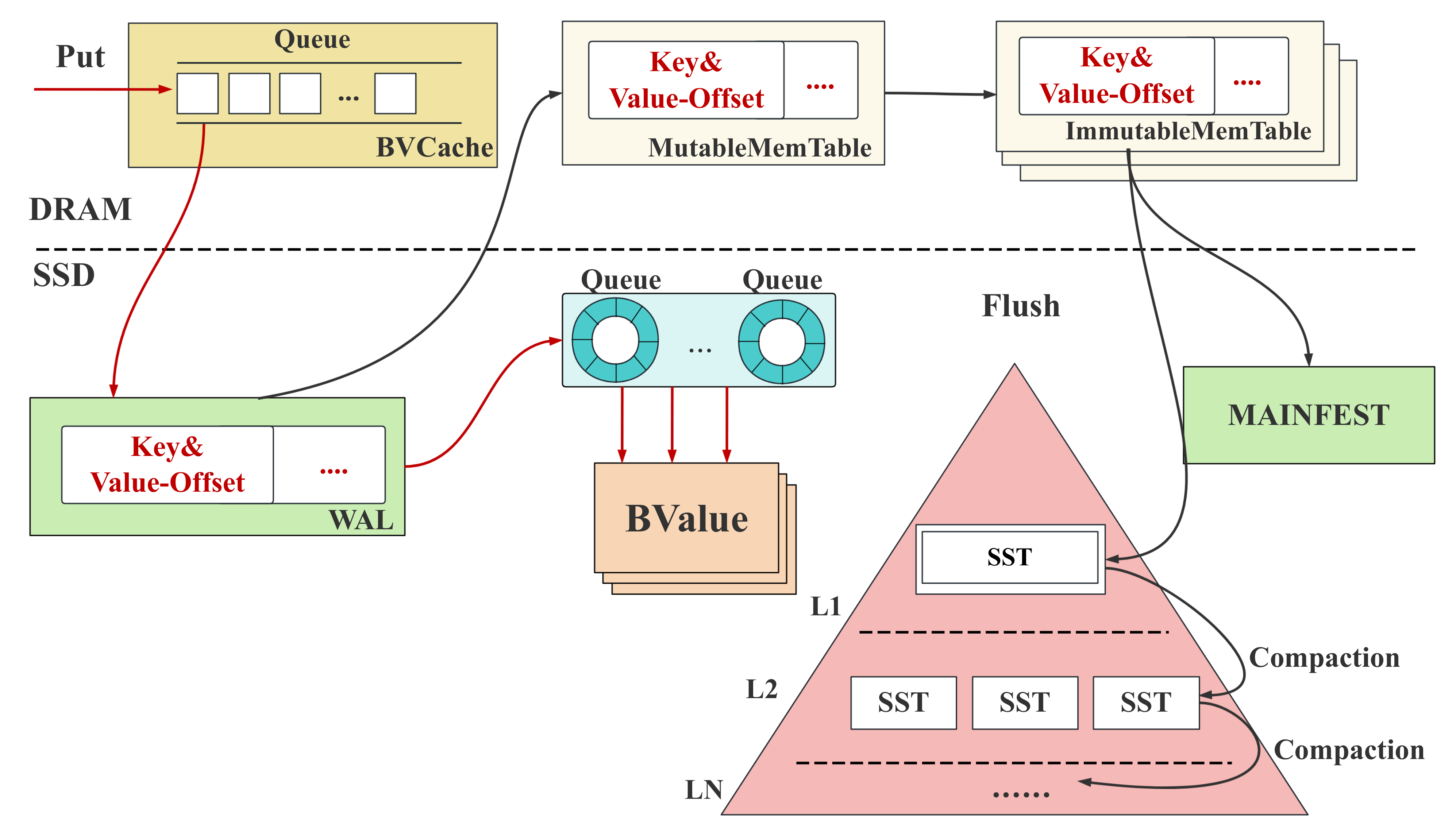}
    \caption{Overview Architecture of BVLSM}
    \label{fig:architecture}
\end{figure}
\subsection{Architecture Overview}
Figure~\ref{fig:architecture} illustrates the overall write process of BVLSM. We introduce two key components: the \emph{BValue} file and the BVCache. The \emph{BValue} file is designed to support high-throughput write operations using NVMe SSDs with multi-queue parallelism. It serves as a dedicated storage location for big values, decoupled from the LSM-Tree standard \emph{SSTable} pipeline. In parallel, BVCache is a read-optimized in-memory buffer that temporarily holds recently written hot values. By exploiting temporal and spatial locality, it reduces redundant reads from disk-resident \emph{SSTables} and mitigates read amplification.

BVLSM processes PUT requests through an early-stage key-value separation mechanism. When a request arrives, the system first checks whether \emph{value} exceeds a predefined size threshold. If so, the key-value pair is split during the WAL stage: the big value is redirected to the \emph{BValue} file, and its associated metadata, including file offset and size, is encapsulated as a \emph{ValueOffset} object. The original entry is then transformed into a lightweight \emph{Key-ValueOffset} structure. This compact metadata is written to the WAL log for durability, inserted into \emph{MemTable}, and eventually flushed into \emph{SSTable}. This design significantly reduces write amplification and memory pressure while maintaining compatibility with LSM-Tree semantics.

The core design of BVLSM is detailed as follows:

\begin{itemize}
    \item \textbf{Early-stage key-value separation:} By performing key-value separation before WAL logging, the system significantly reduces memory pressure in \emph{MemTable}, decreases the frequency of flush operations, and minimizes WAL write volume. This early separation improves write efficiency, especially under big-value or high-throughput workloads, by avoiding redundant in-memory buffering and reducing downstream I/O overhead.
    
    \item \textbf{Multi-queue parallel store for big value:} To handle the stream of separated big values, BVLSM leverages the native multi-queue parallel write capabilities of NVMe SSD. Each \emph{BValue} file is assigned to a dedicated write queue, and the system distributes big value data across multiple queues based on an internal scheduling strategy. This design maximizes per-queue write bandwidth, minimizes contention on individual queues, and significantly improves overall write throughput.
    
    \item \textbf{Big value cache for read optimization:} To mitigate read amplification and latency introduced by extra access to \emph{BValue} files during key-value separation, BVLSM employs a fixed-size in-memory buffer that caches recently written key-value pairs. This buffer temporarily holds values before they are persisted to \emph{BValue} files, and maintains a cache of hot data using a replacement policy to ensure fast access to frequently used entries.
\end{itemize}

\subsection{Early-stage key-value separation}

\begin{figure}[t]
    \centering
    \includegraphics[width=1.0\linewidth]{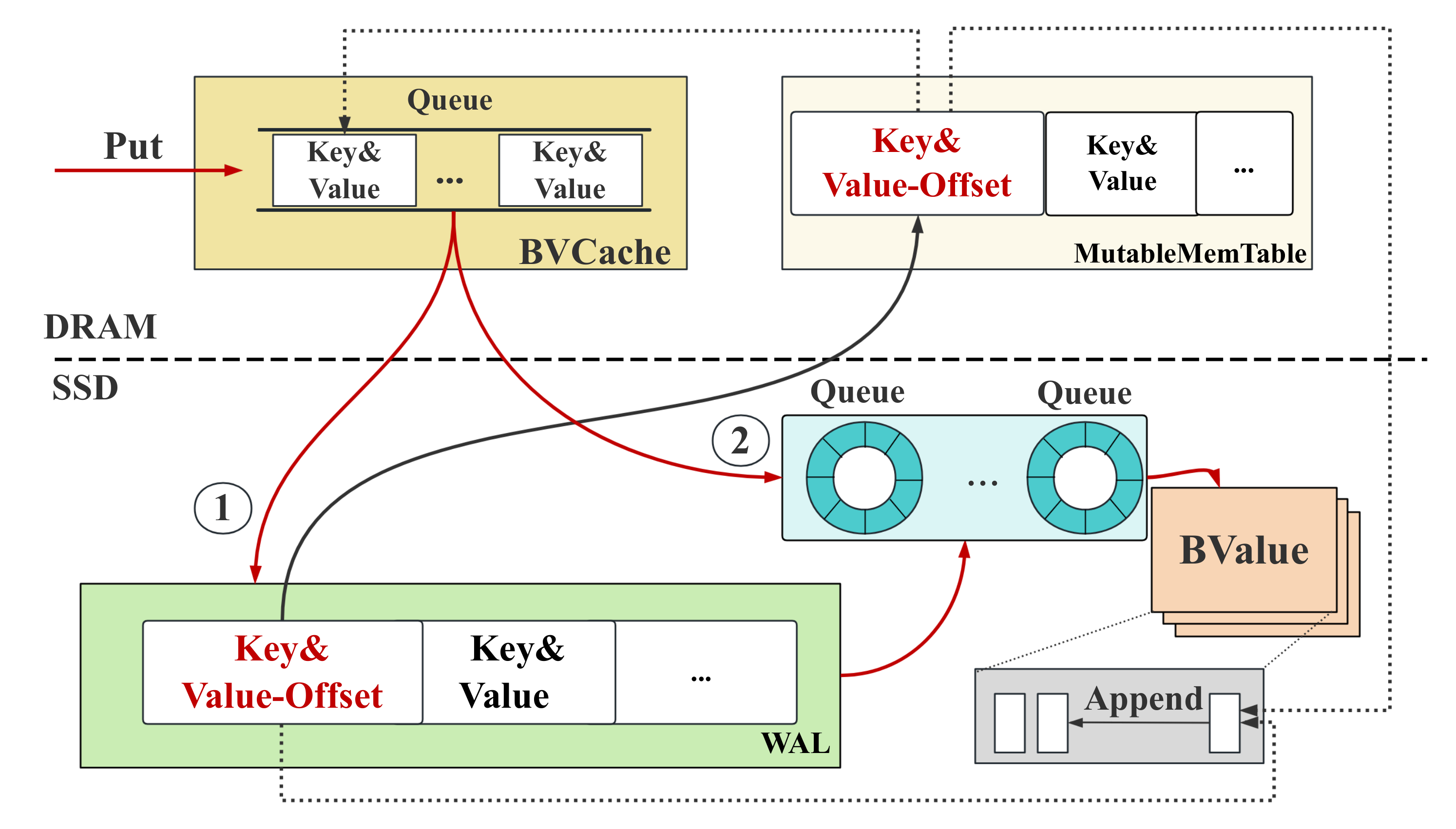}
    \caption{Early-stage key-value separation of BVLSM}
    \label{fig:KV Separation}
\end{figure}

As shown in Figure~\ref{fig:KV Separation}, the core optimization goal of BVLSM is to improve the memory utilization of \emph{MemTable} and reduce the amount of data written to the WAL by implementing a threshold-based big-value separation strategy. The specific implementation path is as follows: values exceeding a preset size threshold are separated from the \emph{MemTable} and subsequent \emph{SSTable} storage paths of the LSM-Tree. During the write phase, they are stored separately in dedicated \emph{BValue} files, while only the \emph{Key-ValueOffset} structure containing the location metadata is retained in the memory index. This hierarchical storage design significantly reduces the memory caching requirements for big values and increases the effective capacity of \emph{MemTable} by approximately N times (the exact factor depends on the separation threshold and the data distribution). In addition, it reduces unnecessary migration of big-value data during subsequent compaction.

In contrast, small \emph{value} objects are written directly into contiguous \emph{MemTable} memory and persistently stored alongside their \emph{Key}s in the \emph{SSTable}s. This hybrid approach balances memory efficiency and I/O overhead: for \emph{value}s smaller than the threshold (e.g., 4~KB), the spatial locality of co-located storage offsets the metadata overhead of separation and avoids serialization or I/O granularity issues associated with small object isolation. By tuning the separation threshold, the system effectively balances memory usage and access performance, achieving improved write throughput and reduced memory footprint in practice.

In the write architecture of the LSM-Tree, the use of a WAL is not mandatory, and its activation is a configurable option for users. Based on this, BVLSM's data writing mode can be categorized into two types:

\begin{enumerate}
    \item \textbf{WAL Enabled Mode (Strong Consistency Guarantee)}: In this mode, the value data are first written synchronously(fsync) to the \emph{BValue} file, ensuring that it persists on disk. Then, metadata are generated, and the system calculates the offset of the value within the \emph{BValue} file, records the value's size (ValueSize), and the \emph{BValue} file path (FilePath). In memory, a new KV record, consisting of the \emph{Key-ValueOffset} pair, is synchronously written to the WAL file. Once the write is successfully committed, the updated KV pair is written to the \emph{MemTable} and subsequently to the \emph{SSTable}. This strategy guarantees that even in the event of a crash recovery, the system can correctly locate and restore the value data, adhering to the strict consistency requirements of the WAL mechanism.
    
    \item \textbf{WAL Disabled Mode}: In this mode, the system adopts an asynchronous write strategy to improve write performance. The value is first stored in an in-memory buffer, avoiding the overhead of frequent disk writes. The system then asynchronously writes the values to the \emph{BValue} file, with a background thread periodically batch-writing the values to the \emph{BValue} file. This method takes advantage of the parallel writing capabilities of NVMe SSDs to enhance throughput.
\end{enumerate}

\subsection{Multi-queue Parallel Store for Big Value}

To fully leverage the multi-queue parallel write capabilities of NVMe SSD and improve the storage efficiency for big \texttt{value} objects, BVLSM introduces a sequential-write-optimized \emph{BValue} file, also known as the Value Log. By decoupling big-value data from key-value pairs and writing them sequentially to \emph{BValue} files, the system not only maintains consistency guarantees but also optimizes SSD write performance.

The \emph{BValue} mechanism consists of two key components: file structure, write strategy.

\emph{BValue} File Structure: Each value is appended sequentially to the \emph{BValue} file, generating a globally unique logical offset that serves as a location identifier. Metadata including the offset, value size (in bytes), and the corresponding \emph{BValue} file path is recorded. This metadata is maintained by both the \emph{MemTable} and the LSM-Tree  \emph{SSTables}. During queries, the system uses the key to quickly locate the physical position of the value, avoiding performance degradation during compaction caused by storing big values directly in the LSM-Tree.

Write Strategy: To fully exploit the parallelism of multi-queue SSDs, the system adopts a multi-\emph{BValue} file and multi-queue binding architecture. big value data is distributed across \emph{BValue} files based on a load-balancing strategy, maximizing storage efficiency. When a write request is sent, the SSD I/O scheduler selects the target \emph{BValue} file using a hash or round-robin algorithm. The write command is then encapsulated in a submission queue (SQ) conforming to the NVMe protocol, with each SQ associated with a specific \emph{BValue} file. The SSD controller dynamically retrieves commands from all active SQs and dispatches them across flash channels using an adaptive scheduling strategy based on real-time channel load. After data are written, the controller writes the operation status to the corresponding completion queue (CQ), triggering an interrupt to notify the host of the successful write.

The multi-queue design mitigates the performance bottleneck caused by single-queue overload and, when combined with asynchronous batch submission, aggregates small-to-medium-sized value writes. These are aligned to the physical page size before being flushed in full pages, improving parallel write throughput. This design enables close coordination between SSD multi-queue hardware and software-level queue scheduling, allowing for near-linear scalability in storage bandwidth.

\subsection{Big Value Cache for Read Optimization}

In BVLSM, once key-value (KV) separation is performed, value data is stored in dedicated \emph{BValue} files. Without separation, querying a KV pair typically requires multiple disk accesses to read several \emph{SSTables} from the LSM-Tree. After separation, the LSM-Tree shrinks in size, reducing the average number of \emph{SSTable} levels accessed during point lookups from five to two or three, which helps lower read latency. Moreover, because the newly constructed KV pairs (\emph{Key-ValueOffset}) are significantly smaller, a dedicated cache can be built to store frequently accessed \emph{Key-ValueOffset} entries. This further reduces disk I/O by minimizing the number of \emph{SSTables} accessed from the LSM-Tree. Combined with a value buffer that caches recently written values, this mechanism improves access speed and reduces read latency from \emph{BValue} files.

The \emph{BVCache} is a fixed-size in-memory structure with a capacity equal to that of \emph{MemTable}. It caches hot and recently written \emph{ValueOffset} entries by exploiting temporal and spatial locality to reduce disk I/O. Implemented as a double-ended queue, each element stores a \emph{Key}, its corresponding \emph{ValueOffset}, access frequency, and a timestamp. A hash table is used for fast lookup of array indices based on keys.

\begin{figure}[t]
    \centering
    \includegraphics[width=1.0\linewidth]{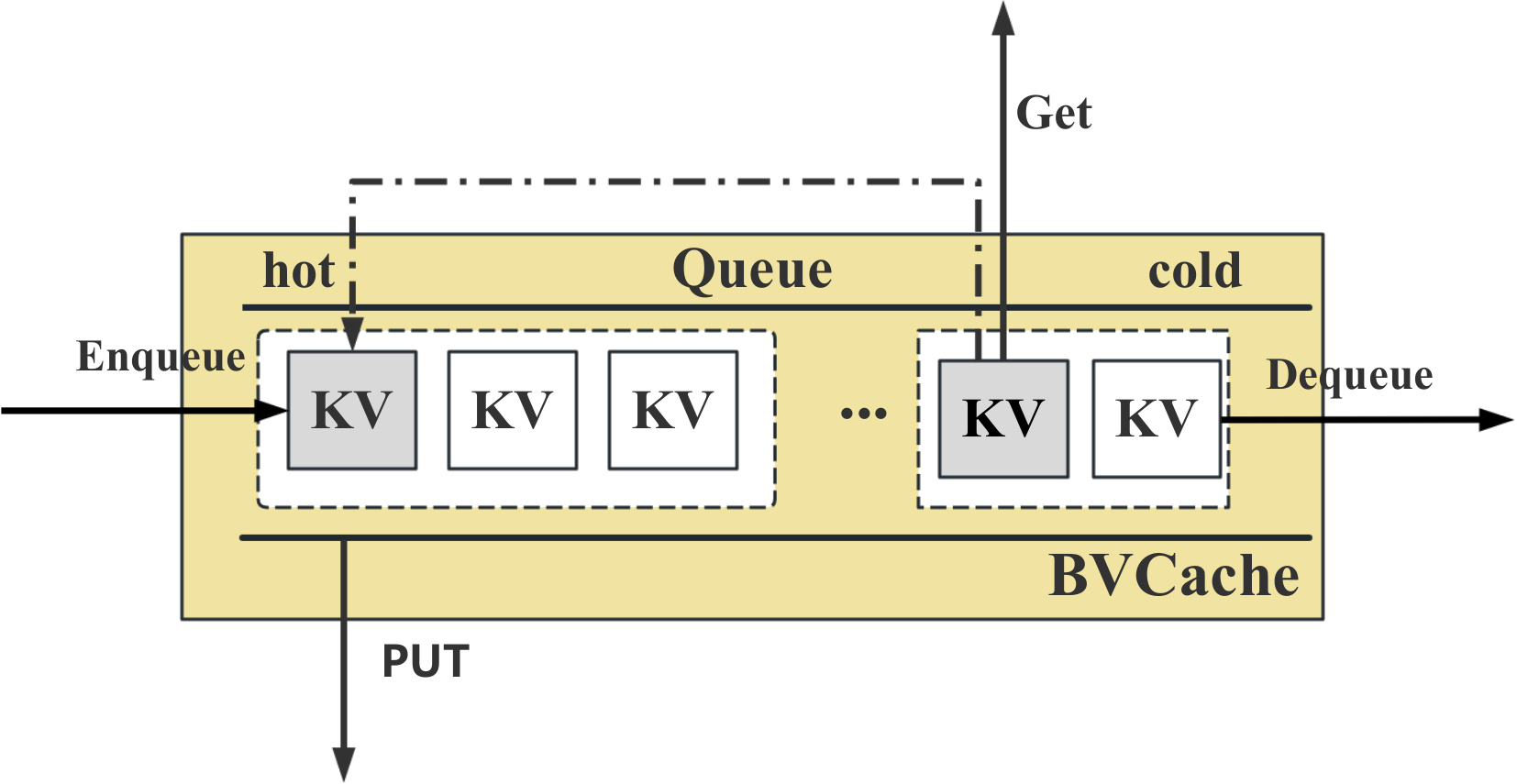}
    \caption{BVCache structure of BVLSM}
    \label{fig:cache}
\end{figure}

As shown in Figure~\ref{fig:cache}, during writes, new data is first stored in \emph{BValue} files, and the returned \emph{ValueOffset} is inserted into the value buffer. The corresponding \emph{Key-ValueOffset} pair is added to the head of the circular array; If the key already exists, the entry is updated and its access frequency incremented. If the buffer is full, the least-recently-used entry at the tail is evicted. For reads, the key is first searched via the hash table; if found, the corresponding ValueOffset is returned, and its access count and timestamp are updated. The value is then retrieved from the \emph{BValue} file using the offset. When eviction is needed, either a time-based (evict the least recently accessed) or frequency-based (evict the least frequently accessed) replacement policy is applied, depending on system load conditions.

\section{Evaluation}
In this section, we first introduce the experimental setup. Then, we evaluate the performance of BVLSM.

\subsection{Experiment Setup}
\textbf{Setup.} The experiments are conducted on a single-node server, with hardware and software configurations shown in Table~\ref{tab:exp-setup}. BVLSM uses the default \texttt{MemTable} size of 128MB, with 1 immutable and 5 mutable \texttt{MemTables}. 

% A flush is triggered when 5 immutable and 1 mutable MemTable (a total of 6) are accumulated. At that point, the oldest immutable MemTable is immediately flushed to an L0 SST file, releasing the memory space.

\textbf{Methodology.} 
We utilize \texttt{db\_bench}~\cite{cooper2010benchmarking} as the primary benchmarking tool to evaluate overall write performance (see Section~\ref{subsec:overall_write_perf}). The dataset size is fixed at 100\,GB, with a constant key size of 16 bytes and value sizes ranging from 4\,KB to 64\,KB, covering a range from small to big-value scenarios. Both RocksDB~\cite{rocksdb} and BlobDB~\cite{facebook_blobdb} are used as baseline systems and are configured identically to BVLSM for a fair comparison. To assess BVLSM under realistic application scenarios, we adopt the YCSB benchmarking suite (see Section~\ref{subsec:YCSB}). We further evaluate the sustained write performance of BVLSM and investigate the performance gains brought by its \emph{multi-queue parallel store} design (see Sections~\ref{subsec:io_stab} and~\ref{subsec:ssd}).

\begin{table}[t]
\centering
\caption{Experimental Setup}
\begin{tabular}{@{}ll@{}}
\toprule
\textbf{Component} & \textbf{Specification} \\
\midrule
CPU     & Xeon\textsuperscript{\textregistered} Gold 5218 (16C/32T, 2.30\,GHz) \\
Memory  & 4 × 32GB DDR4-2933 RDIMM (Total 128GB) \\
SATA SSDs & Samsung SSD 870 (250GB) \\
NVMe SSDs         & Samsung 990 EVO NVMe SSD (1TB) \\
SAS HDD & Seagate ST1200MM0129 (1.2TB) \\
Database Engine & RocksDB v9.7.3 \\
Operating System & Ubuntu 20.04 \\
Linux Kernel & 4.15 \\
\bottomrule
\end{tabular}
\label{tab:exp-setup}
\end{table}

\subsection{Overall write Performance of BVLSM}
\label{subsec:overall_write_perf}

%Write Patterns and WAL Modes
\begin{table}[h]
\centering
\caption{\textit{db\_bench} workloads}
\label{tab:Workloads_dbbench}
\begin{tabular}{llc}
\toprule
\textbf{ID} &  \textbf{Write Patterns}  & \textbf{WAL Modes} \\
\midrule
R-WO  & Random Write           & WAL Off  \\
R-WA  & Random Write           & WAL Async  \\
R-WS  & Random Write           & WAL Sync \\
S-WO  & Sequential Write       & WAL Off \\
S-WA  & Sequential Write       & WAL Async \\
S-WS  & Sequential Write       & WAL Sync \\
\bottomrule
\end{tabular}
\end{table}

\begin{figure*}[t]
    \centering
    \begin{subfigure}[b]{0.32\textwidth}
        \includegraphics[width=\linewidth]{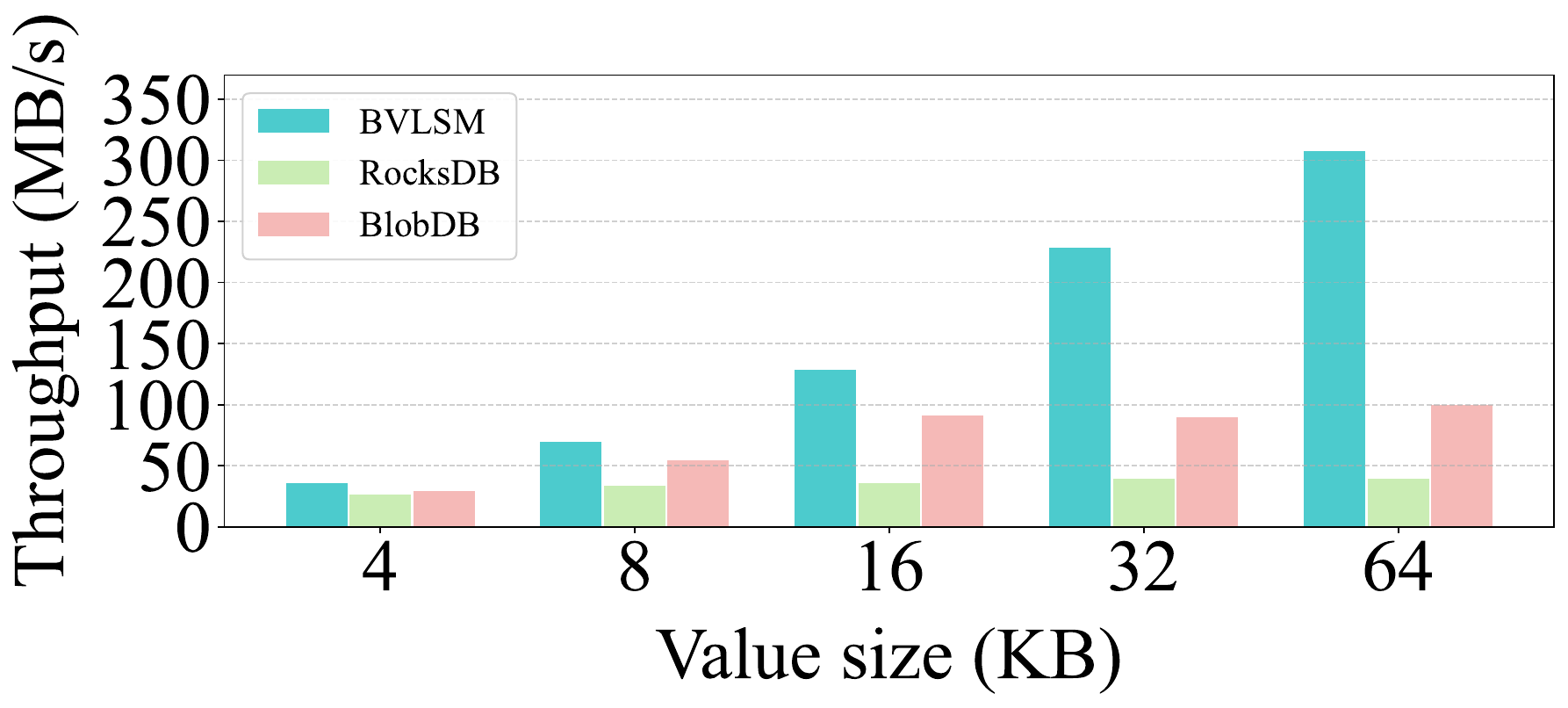}
        \caption{Random write (sync WAL)}
        \label{fig:rand_write_sync_wal}
    \end{subfigure}
    \hfill
    \begin{subfigure}[b]{0.32\textwidth}
        \includegraphics[width=\linewidth]{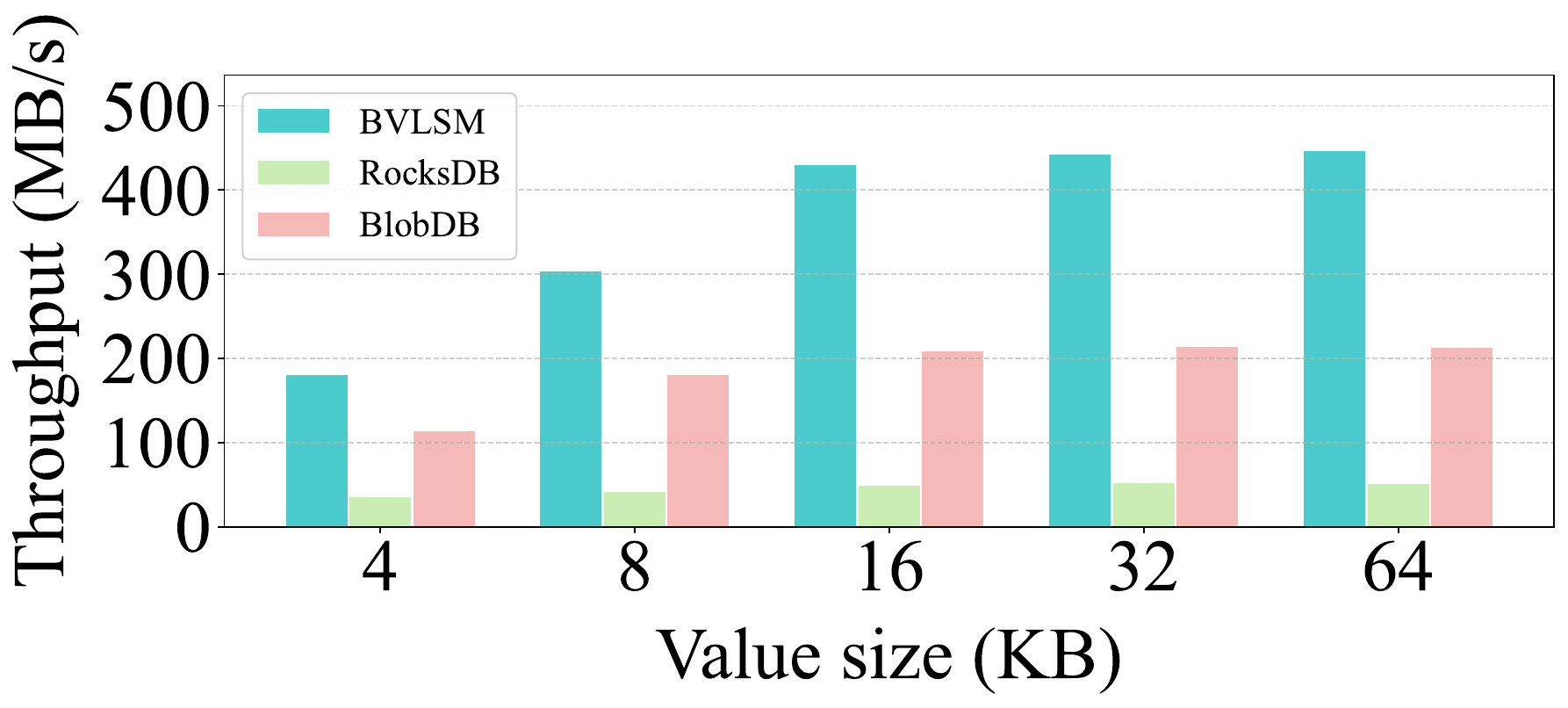}
        \caption{Random write (async WAL)}
        \label{fig:rand_write_async_wal}
    \end{subfigure}
    \hfill
    \begin{subfigure}[b]{0.32\textwidth}
        \includegraphics[width=\linewidth]{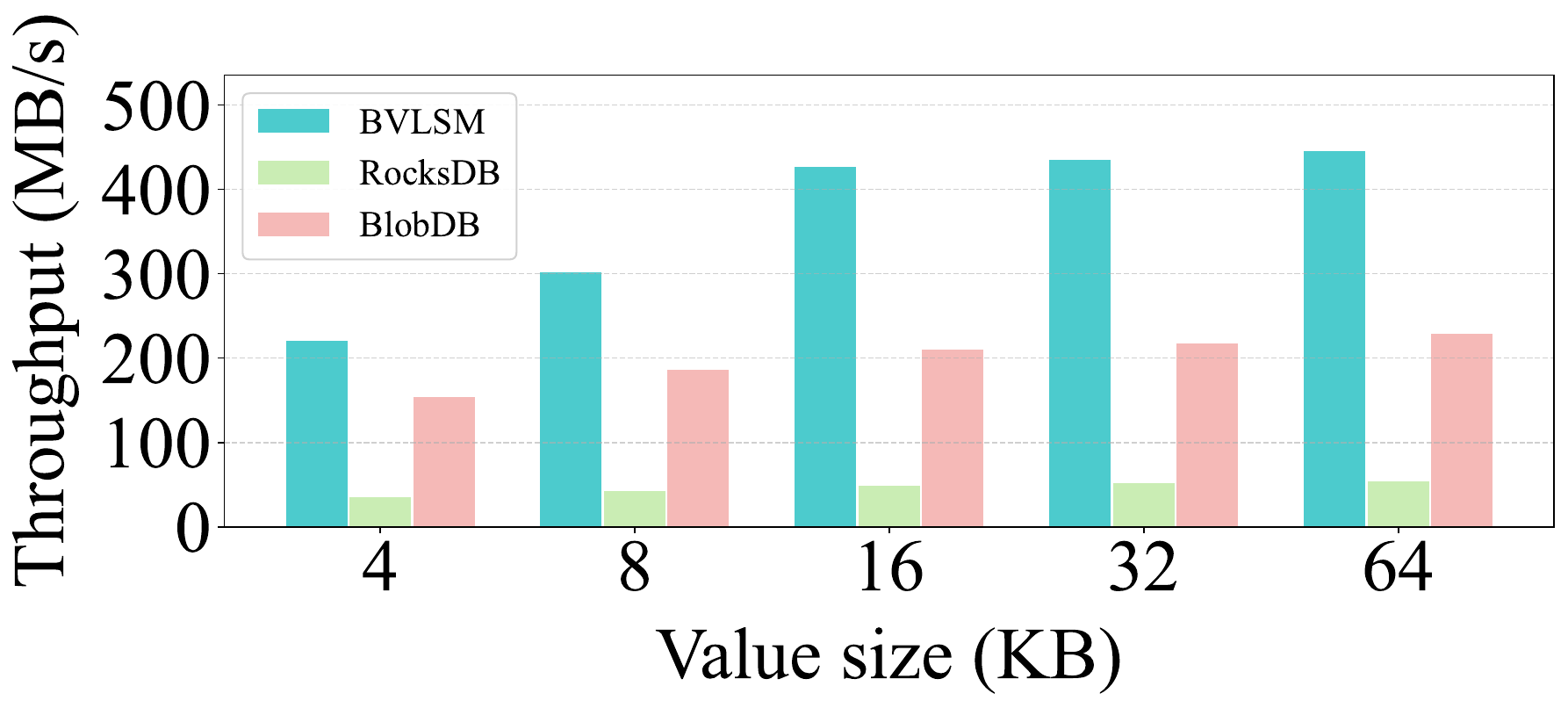}
        \caption{Random write (WAL disabled)}
        \label{fig:rand_write_wal_dis}
    \end{subfigure}
    \caption{Random write performance under different WAL configurations.}
    \label{fig:random_write_overall}
\end{figure*}

\begin{figure*}[t]
    \centering
    \begin{subfigure}[b]{0.32\textwidth}
        \includegraphics[width=\linewidth]{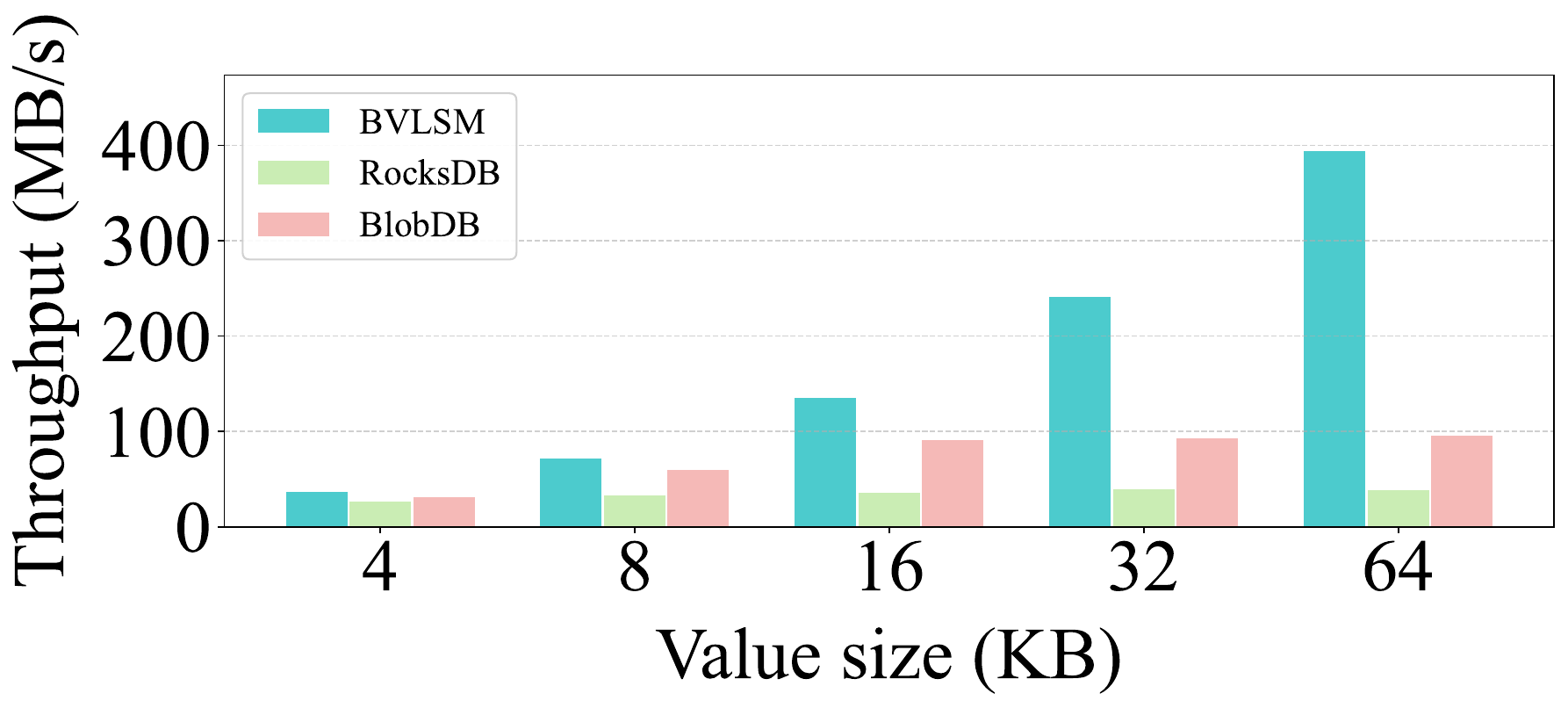}
        \caption{Sequential write (sync WAL)}
        \label{fig:seq_write_sync_wal}
    \end{subfigure}
    \hfill
    \begin{subfigure}[b]{0.32\textwidth}
        \includegraphics[width=\linewidth]{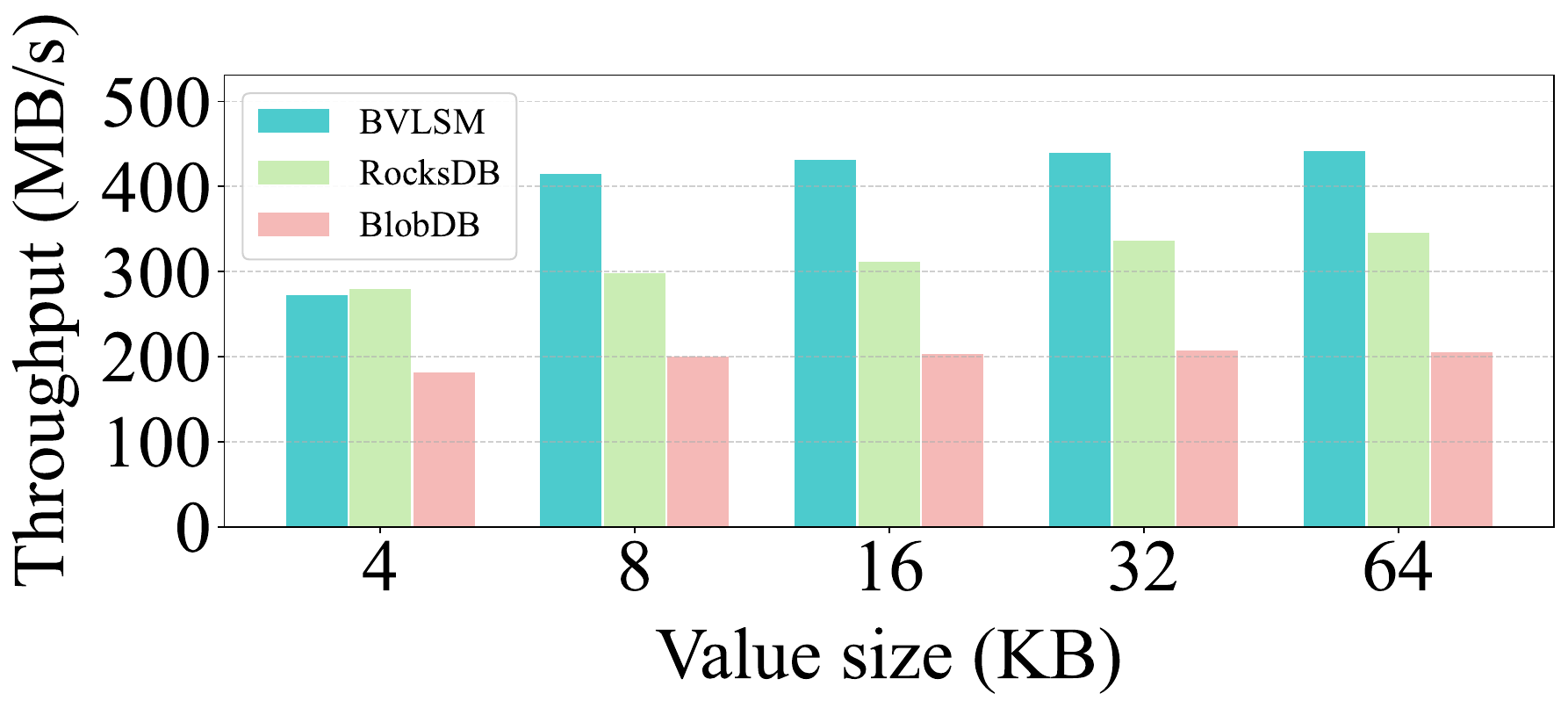}
        \caption{Sequential write (async WAL)}
        \label{fig:seq_write_async_wal}
    \end{subfigure}
    \hfill
    \begin{subfigure}[b]{0.32\textwidth}
        \includegraphics[width=\linewidth]{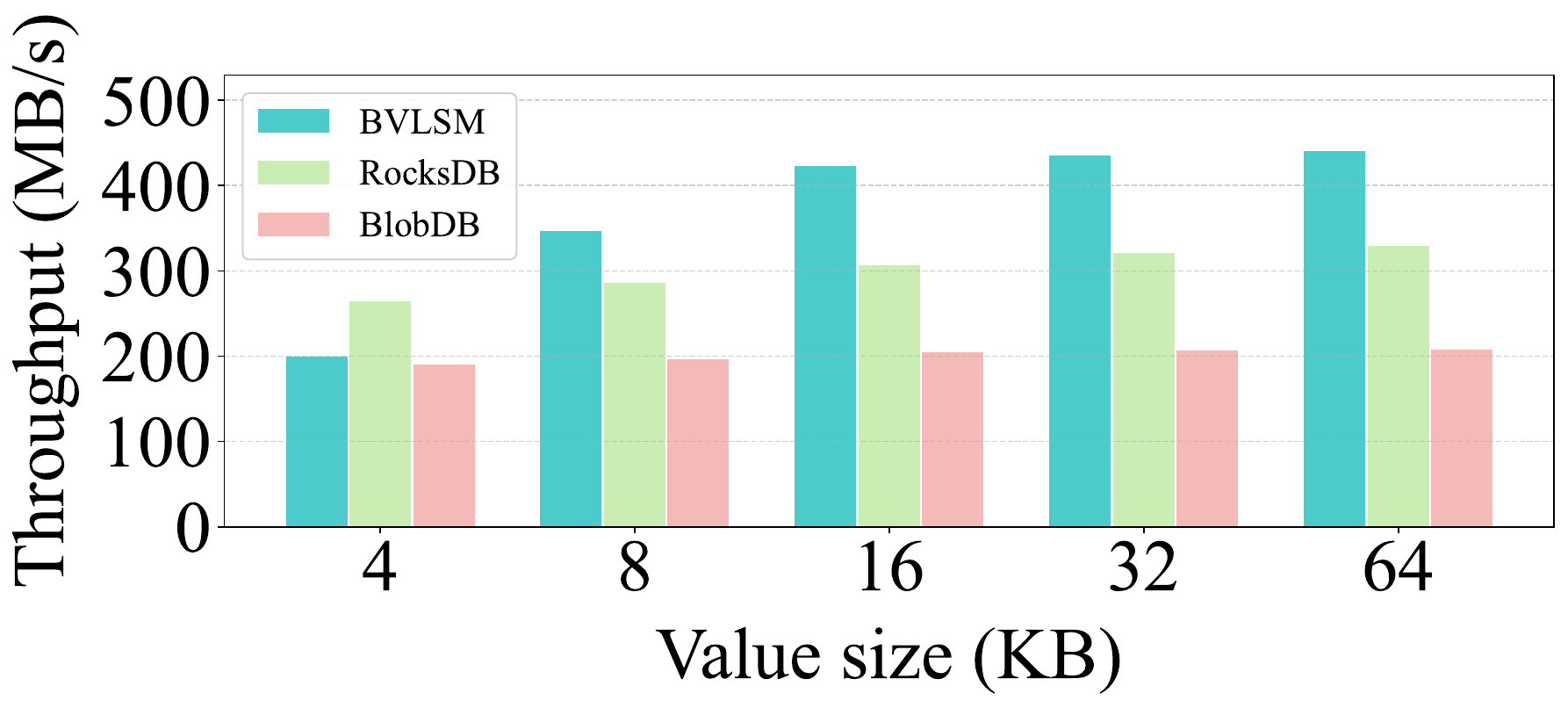}
        \caption{Sequential write (WAL disabled)}
        \label{fig:seq_write_wal_dis}
    \end{subfigure}
    \caption{Sequential write performance under different WAL configurations.}
    \label{fig:sequential_write_overall}
\end{figure*}

We evaluate the write performance of BVLSM in six configurations, combining two access patterns (random and sequential writes) with three WAL modes: synchronous WAL, asynchronous WAL, and WAL disabled. Table 1 summarizes the \texttt{db\_bench} workloads we use. The results are illustrated in two summary figures(Figure~\ref{fig:random_write_overall} for random writes and Figure~\ref{fig:sequential_write_overall} for sequential writes).

As shown in Figure~\ref{fig:random_write_overall}~(a), in workload~R-WS, BVLSM offering approximately 7$\times$ and 3$\times$ speedups over RocksDB and BlobDB, respectively. BVLSM also outperforms both baselines at smaller value sizes such as 4KB, whereas BlobDB shows a significant 57\% drop in throughput at 32KB due to memory pressure and blob file I/O contention. Under workload~R-WA, shown in the Figure~\ref{fig:random_write_overall}~(b), BVLSM outperforms RocksDB and BlobDB by 7.6$\times$ and 1.9$\times$, respectively, in the case of 64KB values. RocksDB performance flattens when value sizes exceed 32KB, while BlobDB throughput increases by less than 2\% beyond this point.

When WAL is disabled~(workload R-WO, as shown in Figure~\ref{fig:random_write_overall}~(c)), the performance trend aligns with that under asynchronous WAL. As value sizes increase, BVLSM exhibits near-linear performance growth and reaches a plateau at 16KB, maintaining high throughput (up to 8$\times$ and 2$\times$ that of RocksDB and BlobDB, respectively, at 64KB). In contrast, both RocksDB and BlobDB show only modest improvements, with RocksDB plateauing after 16KB and BlobDB failing to scale due to internal memory and I/O constraints.

Figure~\ref{fig:sequential_write_overall} further presents the results for sequential writes. In S-WA workload , BVLSM outperforms RocksDB and BlobDB by approximately 1.3$\times$ and 2.2$\times$, respectively. In S-WS workload , the performance advantage becomes even more pronounced, with BVLSM achieving speedups of around 4.1$\times$ over RocksDB and 9.8$\times$ over BlobDB. One observation is that sequential write workloads consistently yield higher throughput than random writes across all systems. This improvement is attributed to the fact that sequential write requests form contiguous memory blocks within the \texttt{MemTable}, which are flushed to the L0 level in batches once the \texttt{MemTable} is full. This batching mechanism reduces the number of I/O operations and improves efficiency.

When WAL is disabled, BVLSM maintains a clear advantage, achieving up to 1.3$\times$ and 2$\times$ higher throughput than RocksDB and BlobDB, respectively, at a value size of 64~KB. While RocksDB exhibits a sharp performance increase for smaller value sizes, which is largely due to its native LSM-Tree structure being well-suited for big-value ingestion. BVLSM continues to demonstrate superior performance for big values, owing to its proactive key-value separation design.

In summary, BVLSM leverages early key-value separation and SSD multi-queue parallelism to reduce write amplification and compaction overhead. This design enable it to consistently outperform RocksDB and BlobDB in both random and sequential writes, particularly with larger value sizes.

\subsection{Latency Performance with YCSB Workload A}
\label{subsec:YCSB}
\begin{figure}[t]
    \centering
    \includegraphics[width=1.0\linewidth]{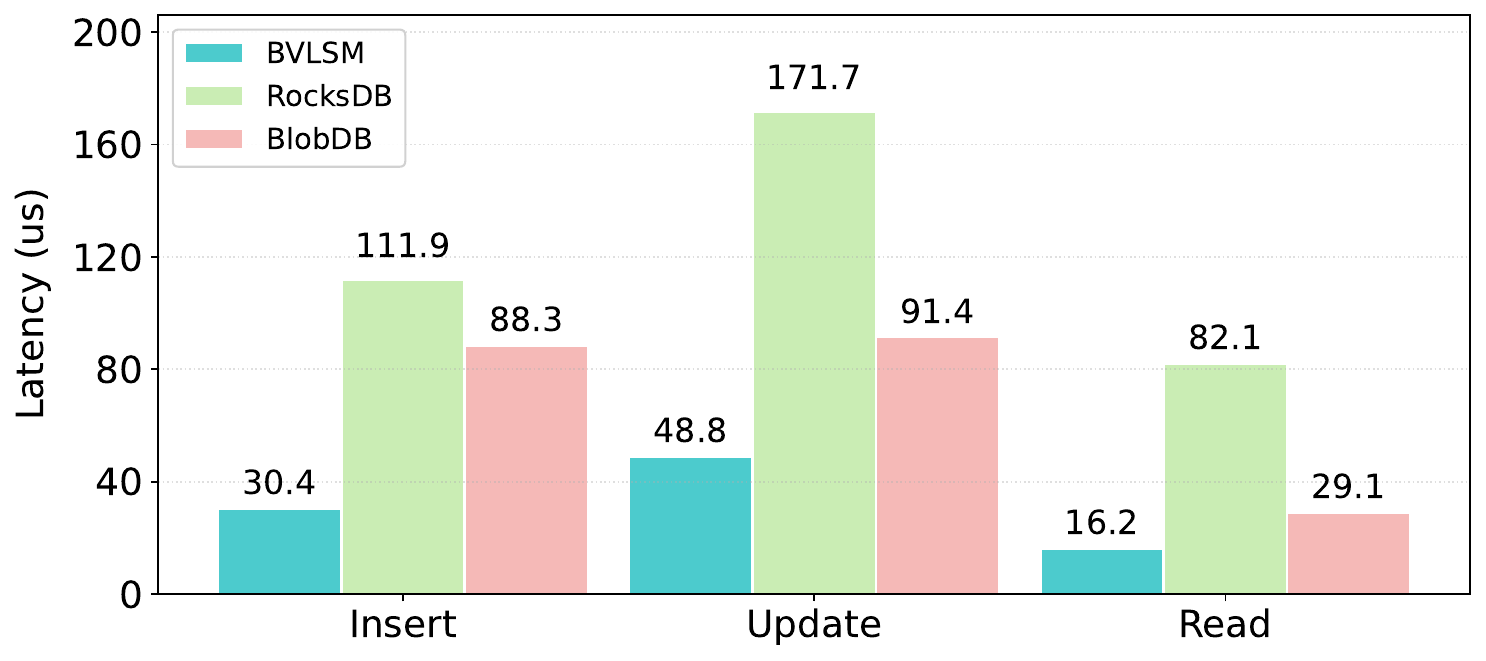}
    \caption{The latency of BVLSM, RocksDB and BlobDB under YCSB-A workload}
    \label{fig:ycsb}
\end{figure}

YCSB is a widely used benchmark that reproduces realistic read--write patterns. To evaluate BVLSM with high-frequency, big-value writes, we select the standard Workload~A, which issues 50\% updates and 50\% reads following a Zipfian key distribution to preserve access locality. To ensure that both \emph{flush} and \emph{compaction} are triggered, we fix the key size at 16\,bytes, the value size at 8\,KB, and preload 500\,000 records. The detailed results are illustrated in Figure~\ref{fig:ycsb}.

RocksDB exhibits a high update latency of 171.7~$\mu$s, highlighting the inherent limitations of traditional LSM-Tree designs in controlling write amplification. BlobDB reduces read latency by 35.5\% compared to RocksDB, but its write performance gains are constrained by its flush-time separation mechanism. In contrast, BVLSM achieves substantial latency reductions across all operations: its insert, update, and read latencies are only 27.2\%, 28.4\%, and 19.7\% of those of RocksDB, respectively. These results demonstrate the effectiveness of the pre-write key-value separation design in mitigating write amplification and improving overall performance.

\subsection{I/O Stability Analysis}
\label{subsec:io_stab}
To evaluate the sustained I/O performance of BVLSM, we use \texttt{db\_bench} to perform a 100~GB random write test with keys of 16~B and values of 4~KB. The throughput of BVLSM, BlobDB, and RocksDB is recorded every 10 seconds to analyze fluctuations in write performance.

\begin{figure}[t]
    \centering
    \includegraphics[width=1.0\linewidth]{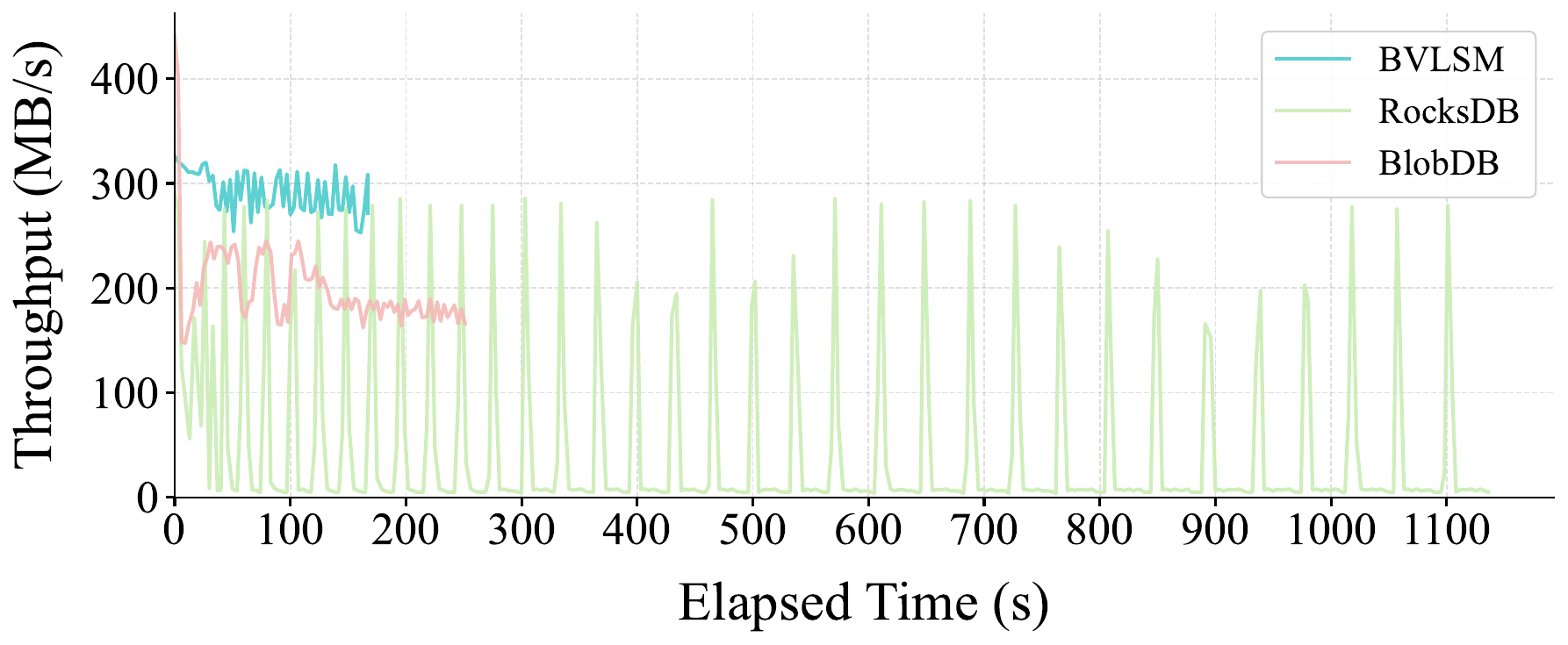}
    \caption{Random write performance of BVLSM, BlobDB, and RocksDB per 10-second interval}
    \label{fig:io_stability}
\end{figure}

As shown in Figure~\ref{fig:io_stability}, the short-term behavior (0--15 seconds) reveals that BlobDB initially achieves high throughput due to fast in-memory absorption. However, this performance rapidly drops to 120~MB/s after 15 seconds, owing to flush queue congestion and memory pressure. In contrast, RocksDB exhibits dramatic throughput oscillations caused by compaction-induced I/O contention within the LSM-Tree structure.

In the long run (after 100 seconds), BVLSM maintains a consistently stable throughput ranging from 250 to 350~MB/s, with the smallest standard deviation among the three systems. On average, it outperforms BlobDB by 2$\times$ and RocksDB by 1.5$\times$, making it better suited for sustained high-throughput workloads. Although RocksDB occasionally peaks, sudden drops at timestamps like 81 and 113 seconds expose intermittent I/O stalls inherent in its architecture, limiting its applicability to latency-sensitive scenarios.

The superior stability of BVLSM is due to its pre-write key-value separation and lightweight metadata design, which reduce flush frequency and compaction triggers. Additionally, its multi-queue BValue writing mechanism balances SSD workload, ensuring sustained bandwidth under high loads. These results validate the robustness of the BVLSM architecture in demanding environments.

\subsection{SSD Mluti-Queue Performance Analysis}
\label{subsec:ssd}
To evaluate the effectiveness of BVLSM's SSD multi-queue design, we conduct experiments using the FIO benchmark on NVMe SSD. Two distinct I/O models are constructed for this evaluation: (i.) Single-queue mode: Four threads pinned to separate physical cores, sharing a single submission queue (SQ). (ii.) Multi-queue mode: Four threads pinned to separate physical cores, each utilizing an independent SQ.

% \begin{itemize}
%     \item \textbf{Single-queue mode}: Four threads pinned to separate physical cores, sharing a single submission queue (SQ).
%     \item \textbf{Multi-queue mode}: Four threads pinned to separate physical cores, each utilizing an independent SQ.
% \end{itemize}

\begin{figure}[t]
    \centering
    \includegraphics[width=1.0\linewidth]{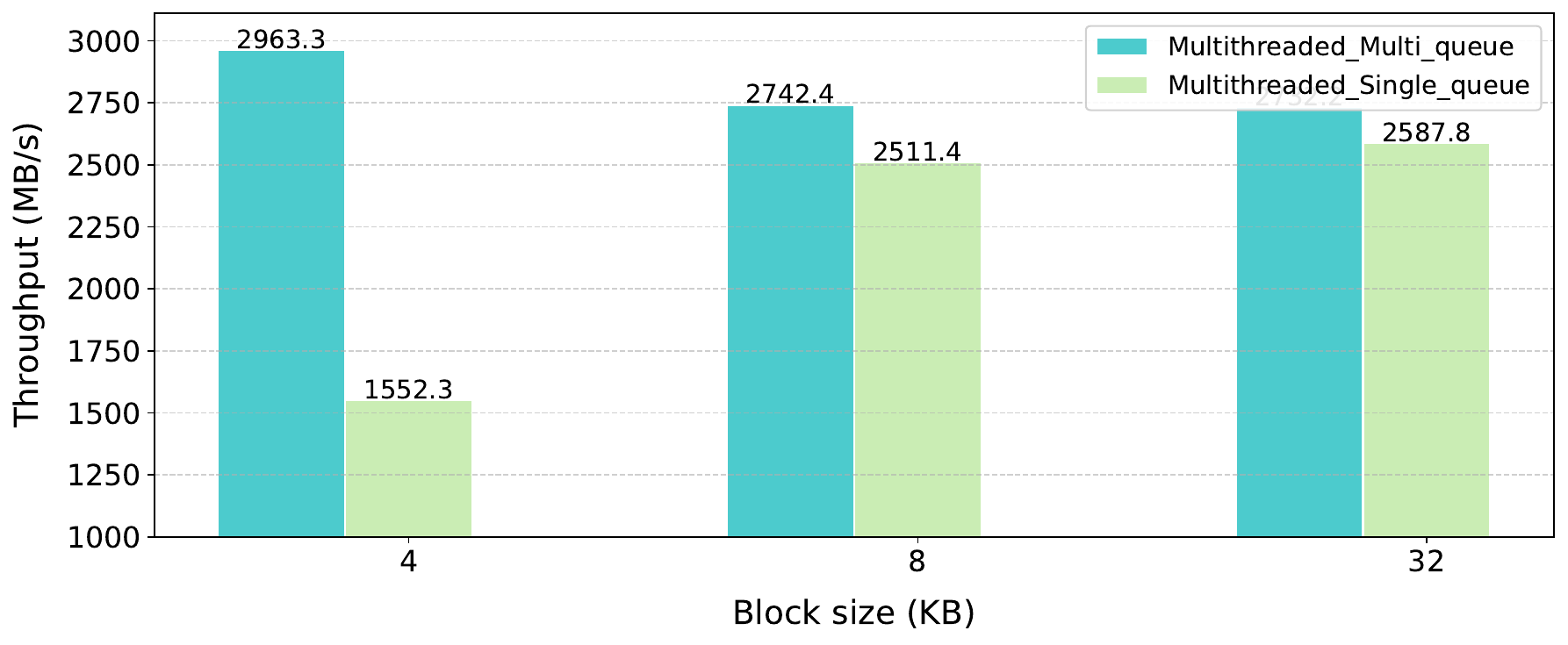}
    \caption{SSD multi-queue and single-queue random write performance with FIO}
    \label{fig:ssd_test}
\end{figure}

As shown in Figure~\ref{fig:ssd_test}, the multi-queue mode delivers a 60.6\% throughput improvement and a 41.2~$\mu$s latency reduction for 4~KB random writes compared to the single-queue setup. This confirms that eliminating lock contention and interrupt bottlenecks significantly benefits high-concurrency I/O workloads. As block sizes increase, the hardware bandwidth gradually becomes the primary bottleneck. However, BVLSM continues to demonstrate strong performance by distributing BValue data across multiple files, achieving up to 2.49$\times$ the BlobDB throughput.

These findings underscore the importance of co-designing key-value separation mechanisms and underlying storage features, such as NVMe multi-queue parallelism, to overcome single-queue bandwidth limits and improve performance in big-value scenarios.

\section{Conclusion}
In this paper, we present BVLSM, a WAL\,--\,time KV separation scheme for LSM-Tree stores targeting big-value workloads. By performing early KV separation during the WAL phase and exploiting SSD multi-queue parallelism together with a big-value cache, BVLSM delivers substantial performance gains. With 64\,KB values under asynchronous WAL, BVLSM achieves $7.6\times$ and $1.9\times$ higher throughput than RocksDB and BlobDB, respectively, while reducing insertion, update, and read latencies to $27.2\%$, $28.4\%$, and $19.7\%$ of those of RocksDB. BVLSM also sustains stable write bandwidth during extended workloads. These results demonstrate BVLSM's ability to improve write performance and stability for big--value scenarios.

\subsection{Acknowledgments}
This work was supported by National Key Research and Development Program of China No.2022YFB2702101, National Natural Science Foundation of China No.92152301.

\bibliographystyle{unsrt}
\bibliography{paper}

\end{document}